\documentclass[letterpaper, 10 pt, hidelinks, conference]{ieeeconf}
\pdfoutput=1
\IEEEoverridecommandlockouts
\usepackage{cite}
\usepackage{amsmath,amssymb,amsfonts}
\usepackage{algorithmic}
\usepackage{graphicx}
\usepackage{textcomp}
\usepackage{hyperref}
\usepackage{comment}
\usepackage{xcolor}
\usepackage{float}

\title{\LARGE \bf SPICE: Smart Projection Interface for Cooking Enhancement}
\author{Vera Prohaska$^{1}$, Eduardo Castell{\'{o}} Ferrer$^{1,2}$\\
\small{$^1$CyPhy Life, Robotics \& AI Lab, School of Science \& Technology, IE University, Spain}\\
\small{$^2$MIT Connection Science, Massachusetts Institute of Technology, Cambridge, USA}}

\begin{document}
\maketitle
\urlstyle{same}
\thispagestyle{empty}
\pagestyle{empty}
{\let\thefootnote\relax\footnote{{Corresponding author: Eduardo Castell{\'{o}}, eduardo.castello@ie.edu}}}

\begin{abstract}
Tangible User Interfaces (TUI) for human--computer interaction (HCI) provide the user with physical representations of digital information with the aim to overcome the limitations of screen-based interfaces. Although many compelling demonstrations of TUIs exist in the literature, there is a lack of research on TUIs intended for daily two-handed tasks and processes, such as cooking. In response to this gap, we propose SPICE (Smart Projection Interface for Cooking Enhancement). SPICE investigates TUIs in a kitchen setting, aiming to transform the recipe following experience from simply text-based to tangibly interactive. SPICE uses a tracking system, an agent-based simulation software, and vision large language models to create and interpret a kitchen environment where recipe information is projected directly onto the cooking surface. We conducted comparative usability and a validation studies of SPICE, with 30 participants.  The results show that participants using SPICE completed the recipe with far less stops and in a substantially shorter time. Despite this, participants self-reported negligible change in feelings of difficulty, which is a direction for future research. Overall, the SPICE project demonstrates the potential of using TUIs to improve everyday activities, paving the way for future research in HCI and new computing interfaces.
\end{abstract}

\section{Introduction}
\label{sec:intro}

The proliferation of ubiquitous computing has transformed everyday objects into interconnected Cyber-Physical Systems (CPS)~\cite{iso_cyber_physical}, where the internet serves as a communication medium between the physical and digital worlds. The field of ``ubiquitous computing'' aimed for a seamless integration of digital technology into our daily lives; rendering technology ``invisible''. However, the dominance of the Graphical User Interfaces (GUIs) has introduced a disconnect between our physical and digital interactions~\cite{m_weiser}.

In the last decades, the adoption of touchscreens and mobile devices introduced new interaction modalities for HCI: touch, gestures, and voice act as core elements of user experience~\cite{iphone2007, ipad2010, ds-gesture-study}. For instance, many of the current mixed-modality interfaces are enhanced by Computer Vision (CV) and Natural Language Processing (NLP), which allows developers to leverage voice and gesture recognition, human-emotion detection, and natural language generation to enhance HCI when interacting with the digital world~\cite{sign-lang-rec, emotion-music, whisper, apple-vision-pro}. These methods departed markedly from the confines of desktop computing (e.g., keyboards, mouse-driven interfaces), and pushed the field of HCI into exploring different methods of user interaction to merge the physical and digital space for users. 

Tangible User Interfaces (TUI) are a possible pathway to move beyond GUIs, as they provide a natural interface to the digital world by providing physical representations of digital information. TUIs provide users with physical as well as visual feedback cues~\cite{tui-ppf}. Originally introduced by Fitzmaurice et. al. as a graspable interface~\cite{fitzmaurice}, TUIs brought a new design paradigm and problem space, as well as an initial definition of interfaces for HCI. For instance, the physical representation of the Windows, Icons, Menus and Pointers interfaces (WIMP) was used with the aim of overcoming the limitations of screen-based interfaces~\cite{Ishii1997, therant}. 

TUIs, like any physical object, operate within specific degrees of freedom (DoF)~\cite{thesis-defense-brygg-ulmer}, leading to the development of abstractions to enhance usability. For instance, Siftables~\cite{siftables}---which consisted of manipulating small, cube-shaped displays that sense their proximity to each other---were used for games, sorting data, or as music interfaces. Further integration of digital information processing into objects enabled TUIs that seemingly merge with our physical space. One such example was \textit{Hermits}, which proposed physical TUIs that could be reconfigured to represent multiple digital entities~\cite{hermits}. Hermits are miniature robots with a differential drive locomotion system and a 3D printed casing giving them a ``shape''. Depending on the task they need to execute, {\it Hermits} can change ``shape'' by going out of the user view and changing the casing. Then, they emerge into the user view with the new casing intact. Recent research in TUIs has applied methods from robotics~\cite{zooids, sui-reactile, tangible-swarms}, providing physical actuation and feedback. Despite these advancements in TUI research, the integration of TUIs into everyday environments, particularly those involving daily physical tasks (e.g., house chroes, cleaning) remains largely unexplored. 

The kitchen, an environment operated by physical processes, presents an opportunity to explore the integration of digital information and functionality. For instance, culinary recipes are typically obtained in the form of text-based information (whether digital or printed), yet preparing a dish still involves a set of techniques that require physical interaction (e.g., stirring), undivided attention (e.g., slicing), and other hands-on skills. This creates a disconnect between the static, inanimate nature of a recipe and the dynamic experience of cooking. While existing research has investigated AR and projection-based solutions for the kitchen~\cite{Ju2001, ar-kitchen-ci, kitchen-ar-kids, ballie, scoping-review-tui-design}, a comprehensive exploration of TUIs and multi-modal input for two-handed interactions in the kitchen is lacking.

\begin{figure*}[t!]
    \centering
    \includegraphics[width=\textwidth]{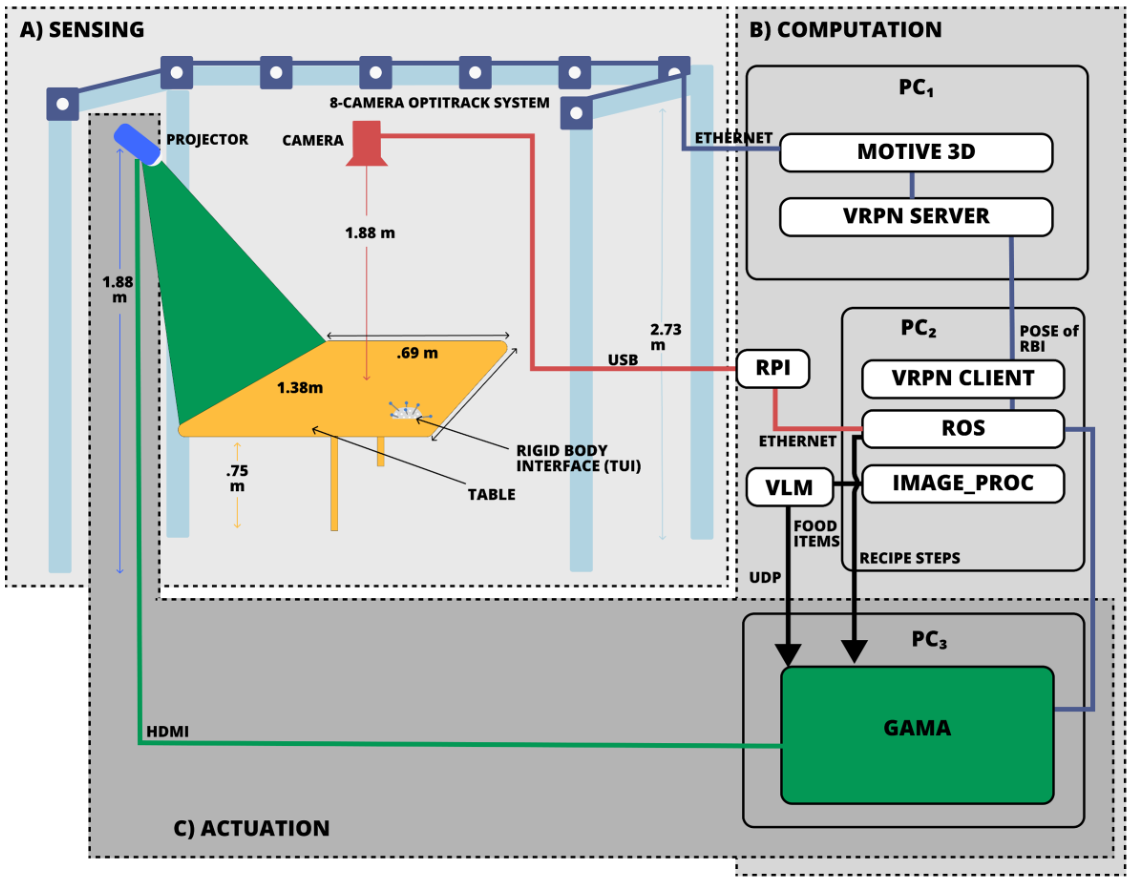}
    \caption{System overview where its three main components are depicted. A) The sensing part of the system includes the use of an optical tracking system (i.e., OptiTrack) to track rigid bodies, a Rigid Body Interface (RBI) (i.e., a 3D-printed base that holds IR reflective markers), and a USB video camera to record the scene where experiments take place. B) The computational element is composed of three PCs. $PC_1$ is running the tracking system processing software (Motive3D), which sends information to $PC_2$ running ROS (Robot Operating System). Finally, $PC_3$ runs GAMA (an agent-based visualization software) that obtains information from $PC_2$ to compute and render the SPICE interface. C) The actuation component includes a short throw projector angled at the working space where the user can interact with the system by moving the RBI.}
    \label{fig:system_overview}
\end{figure*}

This project aims to address this gap by developing SPICE (Smart Projection Interface for Cooking Enhancement); a CPS that leverages TUIs and a multi-modal interface to enhance the cooking experience, offering a more intuitive and engaging interface for interacting with recipes. SPICE utilizes different technologies including an optical tracking system, an agent-based visualization software, and vision language learning models to create a kitchen environment where recipe information can projected onto the cooking space acting as an interactive guide. 

\section{Methods}
\label{sec:methods}

Fig.~\ref{fig:system_overview} shows an overview of the proposed system; where the information flow is also described. In Fig.~\ref{fig:system_overview}, the three main components of the system (sensing, computation, and actuation) are presented. In the following sections we explained each of these components in detail.

\subsection{Sensing}
\label{ssec:sensing}

\begin{figure}
    \centering
    \includegraphics[width=\columnwidth]{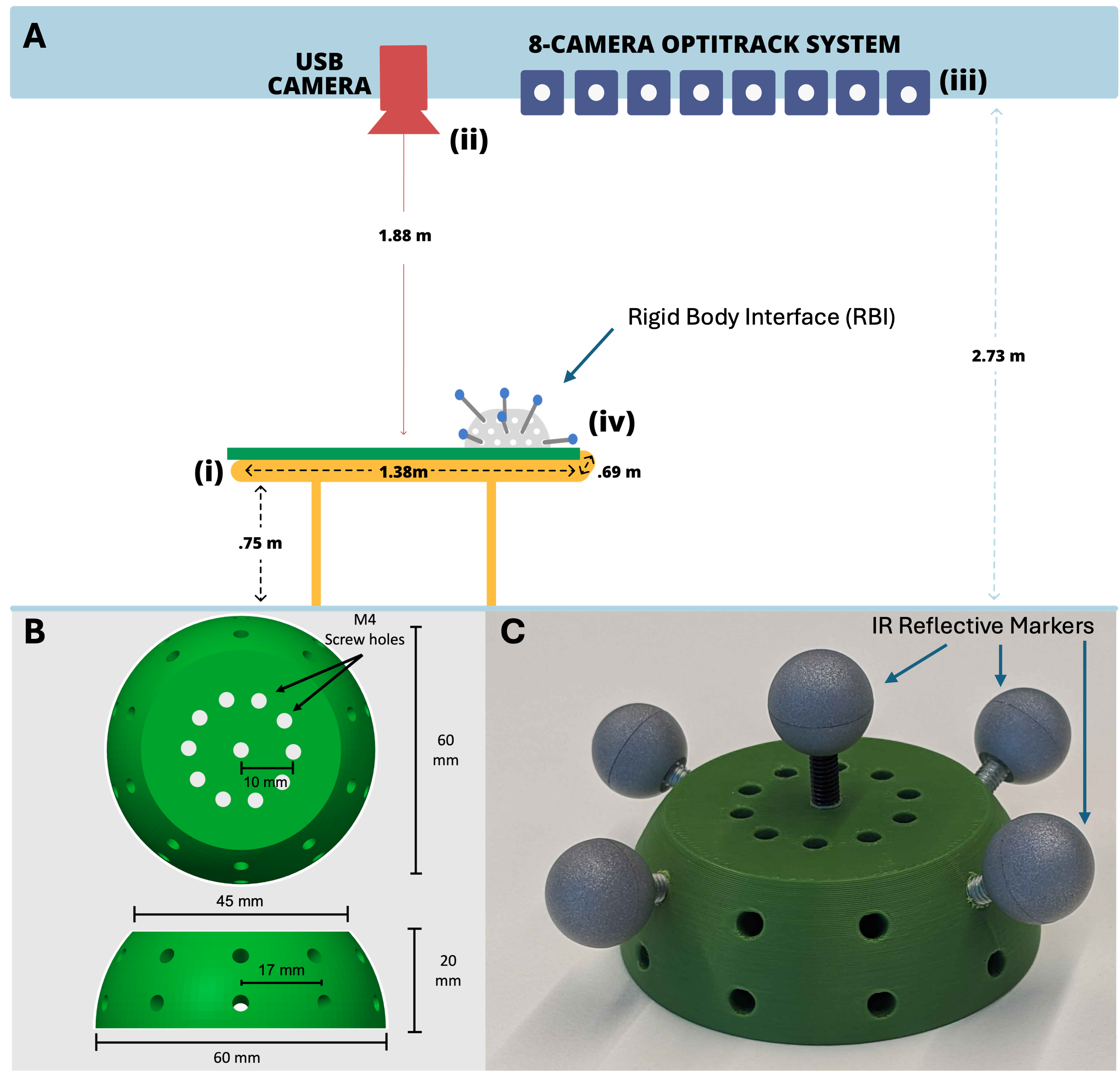}
    \caption{Sensing elements. In A), we can see a representation of the experimental area used in this research. (i) shows the cooking table that acts as the cooking surface. (ii) a USB camera mounted above the table. (iii) An Optitrack system mounted on a truss angled towards the table. (iv) shows a Rigid Body Interface (RBI); a 3D printed piece that holds IR reflective markers detected by (iii). B) shows the CAD model with measures of the RBI. C) depicts one of the RBI after being 3D printed and attaching the retro-reflective markers.}
    \label{fig:sensing_setup}
\end{figure}

Fig.~\ref{fig:sensing_setup} shows the sensing elements of the system. In Fig.~\ref{fig:sensing_setup} A, we can see (i) the table that acts as the cooking surface with the following measures: 1380 (width) $x$ 690 (depth) $x$ 750 (height) mm. Above the table, there is (ii) a USB camera with a 1280x720 pixel resolution. This camera is mounted above (i) and is used to detect the items on the table (e.g., recipe ingredients). In addition, this setup incorporates (iii); an OptiTrack system~\cite{optitrack-system}, to capture and stream the position and orientation of specific objects within the experimental kitchen setup. Finally, (iv) shows a Rigid Body Interface (RBI); a custom-made TUI that holds IR reflective markers and gives the possibility to assemble trackable objects that can be detected by (iii).

In this research, an object is defined by an asymmetric assembly of at least four retro-reflective infrared (IR) markers~\cite{optitrack-docs}. When IR light, emitted by the OptiTrack system (iii), hits a retro-reflective markers, the light is reflected back and captured by the camera's sensor. To accurately reconstruct the 3D position of a marker, at least two cameras are required. This setup allows the system to precisely determine the 3D position (x, y, z) and orientation (roll, yaw, pitch) of an object in the working space of the Optitrack cameras. To hold these markers together, we designed a 3D printable piece (see Fig.~\ref{fig:sensing_setup} B and C) called the Rigid Body Interface (RBI). The RBI has a collection of 31 M4 screw holes to attach IR markers with different heights and therefore with different configurations (see an example in Fig.~\ref{fig:sensing_setup} C). All the files needed to reproduce the RBI are available in the following Github repository\footnote{\url{https://github.com/IERoboticsAILab/3d_printing_designs}}.

\subsection{Computation}
\label{ssec:computation}

Fig.~\ref{fig:system_overview} B) shows SPICE computational pipeline. First, the Optitrack tracking system is connected to its processing software Motive 3D~\cite{motive-software} via Ethernet. This software runs in $PC_{1}$. The Motive 3D software calculates the position and orientation of the RBI based on the process described in section~\ref{ssec:sensing}. Then, Motive3D opens up a Virtual Reality Peripheral Network (VRPN)~\cite{vrpn} server that accepts incoming connections to send the position and orientation data.

$PC_{2}$ receives the position and orientation of RBIs by using a VRPN client. Then, this information is published into the Robot Operating System (ROS)~\cite{ROS} server running locally. We decided to use ROS because it is the most widely used framework which provides ready-made tools to build and connect robotic systems. In addition, image data is also published into ROS from the USB camera (Fig.~\ref{fig:sensing_setup} A (ii)). This data comes from a Raspberry Pi 4B connected to the USB camera, which then sends image data to ROS. Then, the image is rectified by using the $image\_proc$ ROS package in $PC_{2}$. Finally, this data is sent to a Vision Language Model (VLM) to detect the food items (i.e., recipe ingredients) on the table. By utilizing a VLM (gpt-4o) and a large language model (gpt-3.5-instruct), food items can be recognized in the images sent. Once food items are recognized (e.g., tomatoes, onions, avocados, lemon), the system sends them via UDP to the software running the SPICE interface (i.e., GAMA). Finally, the position and orientation of the RBI, are converted into UDP messages using ROS topics and subsequently transmitted to $PC_{3}$.

$PC_{3}$ runs an agent-based visualization software called GAMA~\cite{gama-platform}. GAMA was chosen due to its robustness in processing real-time information, mapping 3D spaces, and agent-based modeling features. One of the main design premises for this research is that every object present in the kitchen environment becomes an agent. An agent as defined in GAMA can have several properties, such as \textit{networking}, \textit{locomotion}, and interaction with other agents~\cite{gama-platform-interaction-agents}. For instance, agents that have networking property can have an IP address. In this regard, objects like the food ingredients, RBIs, etc. can have their own IP which they use to receive and send data in real time. Finally, $PC_{3}$ renders the User Interface (see Fig.~\ref{fig:ui}), and projects it onto the table by using a short throw projector.

\subsection{Actuation}
\label{ssec:actuation}

Fig.~\ref{fig:system_overview} C) shows the actuation component used in this project. This setup includes an Optoma Short throw Projector (UHD35STx) connected to $PC_3$ via an HDMI cable. This projector displays information onto a table set in the middle of the working space. By using a projector angled towards the table, information can be displayed directly onto the cooking surface and on top of objects, following other projector-based visual interaction setups~\cite{Ju2001, zooids, tangible-swarms, super-food}. 

\section{User Interface}
\label{sec:SPICE}

\begin{figure*}
    \centering
    \includegraphics[width=.88\textwidth]{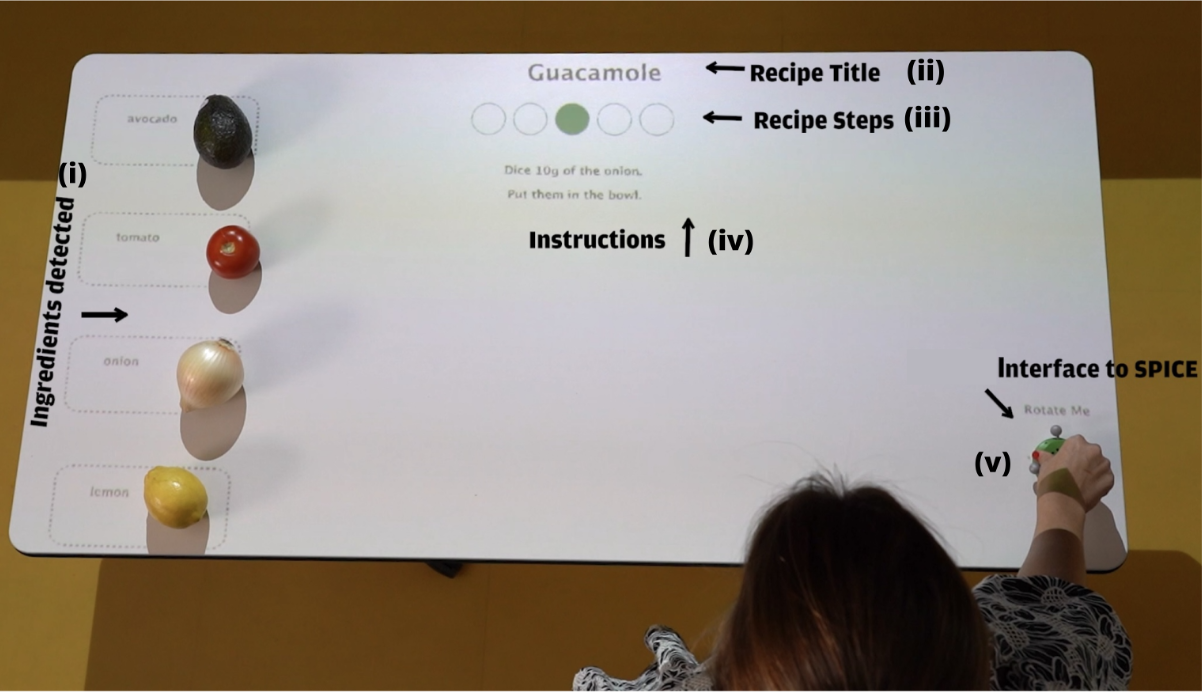}
    \caption{SPICE User Interface. This figure shows the layout of the User Interface (UI) used in the project. From left to right: (i) ingredients detected by the VLM are placed in separated boxes. (ii) The recipe title is displayed together with its corresponding steps (iii). Detailed instructions (iv) such as cooking times and ingredient weights are displayed below the corresponding step. Finally, the RBI (v) is placed at the bottom right of the cooking surface to scroll among the recipe steps. }
    \label{fig:ui}
\end{figure*}

The interaction happens on top of the table, where the table and the projection are considered the ``interface'' (as seen in Fig.~\ref{fig:ui}). This interface works with the following user workflow: 1) The user presents the ingredients for the recipe, and places them ontop of the table. Then, these items are detected by the VLM using the images coming from the USB camera. 2) The labels for these ingredients (e.g., Tomato, Onion, Avocado) are sent to $PC_{3}$ via UDP and rendered into boxes at the left-hand side of the interface where the user places the corresponding ingredients (Fig.~\ref{fig:ui} i). 3) A recipe (e.g., guacamole\footnote{\url{https://en.wikipedia.org/wiki/Guacamole}}) that matches the detected ingredients is presented to the user. Then, its title (Fig.~\ref{fig:ui} ii), steps (Fig.~\ref{fig:ui} iii), and detailed instructions (Fig.~\ref{fig:ui} iv) are shown. Recipe steps are depicted as small ``bubbles'' aligned in as a list. 4) On the bottom right side, the interface provides an area to place the RBI (Fig.~\ref{fig:ui} v). In this sense, the RBI works as a way to interact with the interface. For instance, users can flip through the recipe steps by rotating the RBI (both clockwise and counterclockwise) only within that designated area. A video depicting SPICE can be found in our Youtube Channel\footnote{\url{https://www.youtube.com/watch?v=BouEhriwqQ0}}. In addition, all the code needed to run SPICE is available in the following Github repository\footnote{\url{https://github.com/IERoboticsAILab/SPICE}}.

\begin{figure}[tbh] 
\centering
\includegraphics[width=\columnwidth]{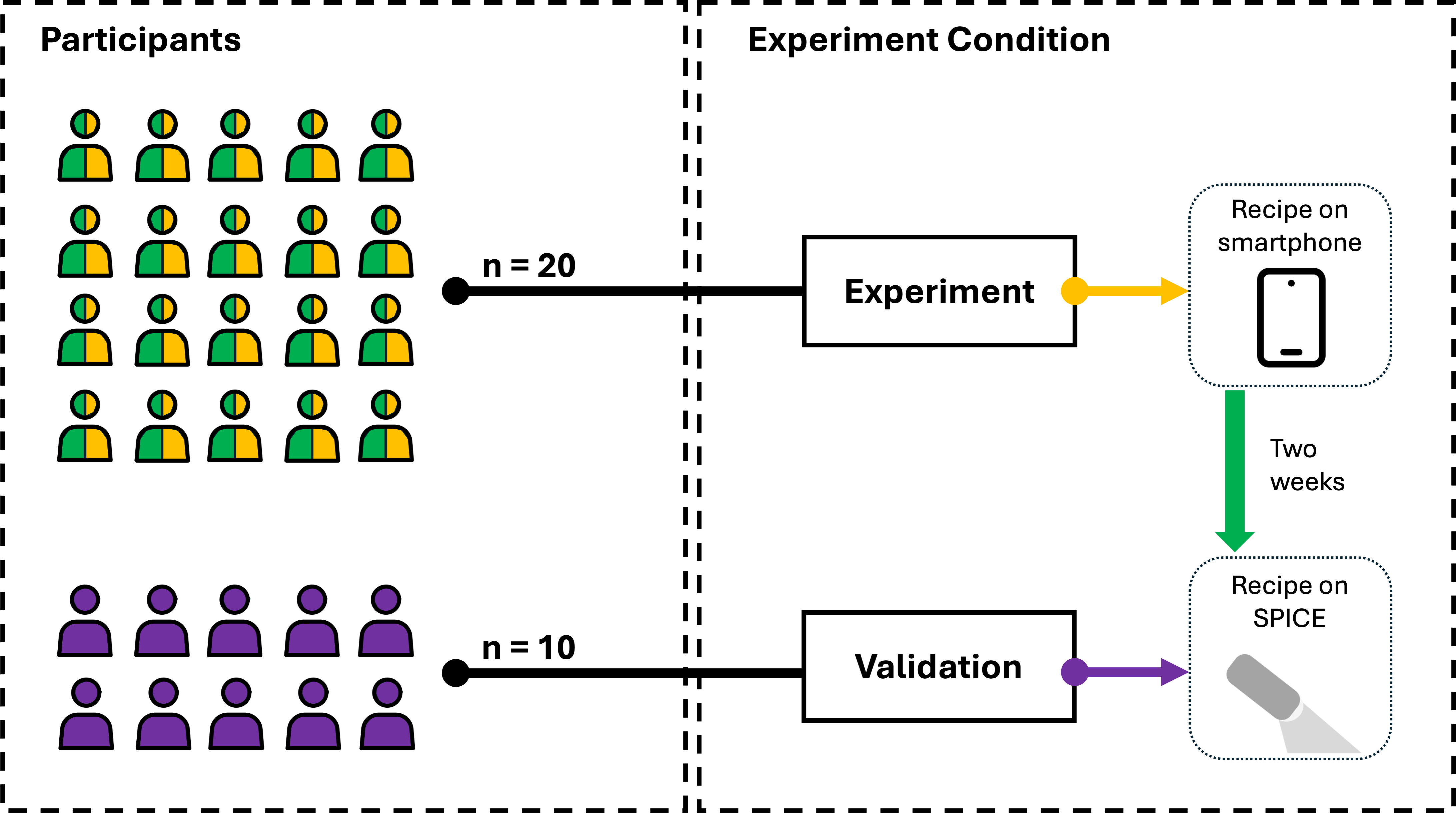} 
\caption{Overview of the experimental design. A first group of 20 participants (Experiment group) were invited first to follow the recipe on the smartphone (orange) and after a two-week interval follow the recipe by using SPICE (green). A second group of 10 participants (Validation group) exclusively used SPICE to follow the recipe (purple).}
\label{fig:group_design}
\end{figure}

\section{Experiment Setup}
\label{sec:experiments}

A comparative usability test was employed to evaluate the effectiveness of SPICE against a traditional recipe-following process (e.g., reading a text-based recipe page from a smartphone). A separate validation test was employed to measure the effect of using SPICE independent of recipe familiarity (which cannot be done with the comparative usability test, due to the inherent limitations of this test type). 

\begin{figure*}[!t] 
    \centering
    \includegraphics[width=0.88\textwidth]{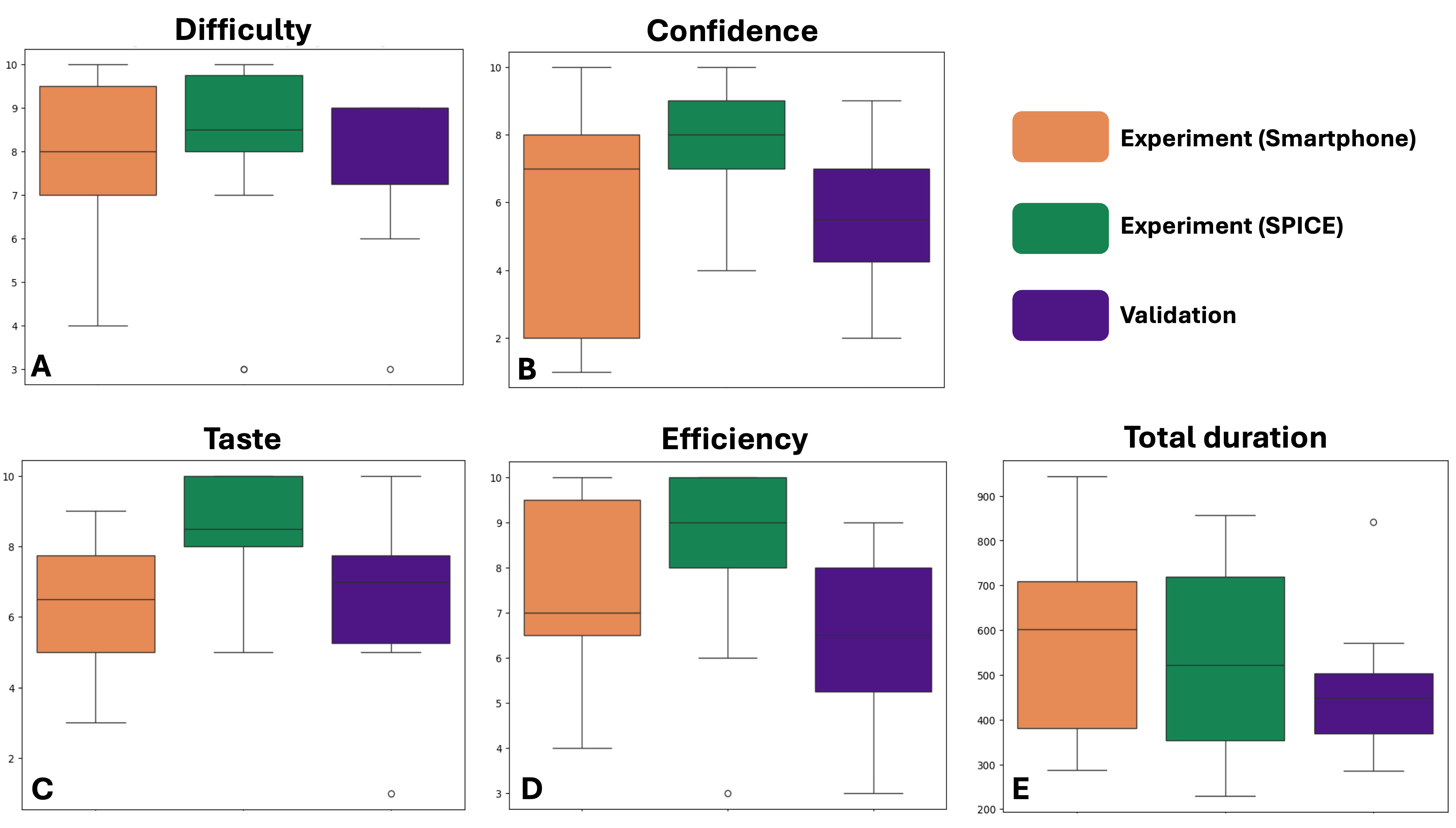}
    \caption{Overview of the results obtained grouped by each one of the metrics considered. A) shows a boxplot with mean and standard deviations of the difficulty perceived for the experiment group that followed the recipe in the smartphone (orange), using SPICE (green), and the validation group (purple). B, C, D, and E show the same information for the Confidence, Taste, Efficiency, and Total Duration (in secs) metrics respectively.}
    \label{fig:results}
\end{figure*}

\begin{table*}[tbh]
\centering
\begin{tabular}{|l|c|c|c|c|}
\hline
\textbf{Metric} & \textbf{Experiment (smartphone) - (A)} & \textbf{Experiment (SPICE)} & \textbf{Validation - (B)} & Percentage Difference \textbf{(A-B)} \\
\hline
Efficiency (1-10) & 7.631 & 8.526 & 6.300 & -17.44\% \\
Confidence (1-10) & 5.310 & 7.895 & 5.400  & +1.69\% \\
Taste (1-10) & 6.166  & 8.444 & 6.500 & +5.41\%\\
Difficulty (1-10) & 7.894 & 8.000 & 7.800 & -1.19\%\\
Total duration (secs) & 559.58 & 524.47 & 471.30 & -15,77\%\\
Number of stops & 10.00 & 6.158 & 6.00 & -40\%\\
\hline
\end{tabular}
\caption{Experiment Results. Comparison of averages across Experiment (smartphone), Experiment (SPICE), and Validation groups for Efficiency, Confidence, Taste, and Difficulty, which are self-reported ratings by the participants, whereas the Total Duration and Number of stops are observed from the video recordings. The last column depicts the percentage difference between the Experiment (smartphone) (A) and the Validation groups (B).}
\label{tab:experiment_summary}
\end{table*}

In the two types of tests, the recipe methods were evaluated according to the following metrics reported by the participants (on a scale from 1 to 10): \textit{Difficulty}, a self-reported rating of recipe difficulty; \textit{Confidence}, a self-reported rating of confidence in own performance in recipe execution; \textit{Taste}, a self-reported rating of the final meal's taste; \textit{Efficiency}, a self-reported rating of efficiency in recipe execution. The user experience of the participants was also quantified using the following two metrics, recorded by the experimenters using video footage: \textit{Total Duration}, the recorded time taken to complete the recipe in seconds, and \textit{Number of Stops}, the recorded number of times a participant stopped cooking to look at the recipe.

The two types of tests were conducted on a 5-step recipe for making guacamole, using four basic ingredients (tomato, avocado, lemon, onion). Participants were provided with a cutting board, a knife, a fork, plates and napkins. Every recipe step required two-handed actions. All participants were video recorded while performing their cooking. After the recipe was completed, all participants filled out a self-evaluating questionnaire\footnote{All questionnaires are available on the GitHub repository of the project.} which tested for the variables of interest using 10-point Likert scales.

As shown in Fig.~\ref{fig:group_design}, 30 participants (ages 20 to 50) were divided into two groups: the Experiment group $(n=20)$ and the Validation group $(n=10)$. The Experiment group followed the guacamole recipe first by using a text-based recipe on a smartphone, then, after a two-week interval, by using the SPICE interface. The Validation group followed the recipe using only the SPICE interface. The Experiment group was composed of 7 women and 13 men, and the Validation group of 1 woman and 9 men. The self-rated level of familiarity with digital technology, which we obtained through participants' responses to the questionnaire (on a scale of 1 to 10, from "not familiar at all" to "extremely familiar"), had an average of 6.17 and a mode of 8 for the Experiment group, and an average of 5.4 and a mode of 8 for the Validation group. Participants were mostly 4th-year undergraduate students, with a few staff members of the university also participating.

Both parametric (Paired Sample T-Test) and non-parametric (Wilcoxon Signed-Rank test) methods were considered for data analysis. The Shapiro-Wilk test was used to assess normality of the data distribution. If non-normality was detected $\alpha < 0.05$, the Wilcoxon Signed-Rank test was chosen due to its suitability for small sample sizes and repeated measures.

\section{Results}
\label{sec:results}

Fig.~\ref{fig:results} and Table~\ref{tab:experiment_summary} present the results of both experiment types: comparative usability test (both on the smartphone and SPICE) and validation test (only on SPICE). 

Participants in the Experiment group reported higher Efficiency (mean: 8.526) and Confidence (mean: 7.895) when using SPICE compared to when using a smartphone (mean Efficiency: 7.631; mean Confidence: 5.31), and compared to the Validation group using SPICE (mean Efficiency: 6.3; mean Confidence: 5.4). The Experiment group also rated the Taste higher (mean: 8.444) when using SPICE compared to using the smartphone (mean: 6.166) and compared to the Validation group using SPICE (mean: 6.5). 

The Duration of cooking tasks was lower in the Experiment group when using SPICE (mean: 524.47 seconds) than when using the smartphone (mean: 559.58 seconds). However, the total duration was lowest in the Validation group using SPICE (mean: 471.3 seconds). 

Similar results were observed for the number of stops: the Experiment group when using SPICE had fewer stops (mean: 6.158) than when using the smartphone (mean: 10), and the Validation group using SPICE had the fewest stops (mean: 6), although only slightly lower than the Experiment group when using SPICE. 

In terms of percentage difference between the Experiment group using the smartphone and the Validation group that was only exposed to SPICE (Table~\ref{tab:experiment_summary}), there was a decrease of 40\% in the Number of Stops and a decrease of 15.77\% in the Duration of cooking. Perceived Difficulty was reduced by 1.19\% while self-perceived Taste and Confidence increased by 5.41\% and 1.69\% respectively. Finally, the self-perceived Efficiency decreased by 17.44\%. 

\section{Discussion}
Compared to using a text-based recipe on a smartphone, SPICE substantially increased the efficiency (lower duration, fewer stops) of cooking from a recipe for first-time users in our tests, both when using SPICE after using a smartphone for the same recipe, and when using SPICE without any prior introduction to the recipe. However, first-time users did not generally feel the efficiency improvement in their subjective experience, although they were more likely to report a feeling of greater efficiency if they used SPICE after having used a smartphone. 

In the comparison between using a smartphone versus using SPICE, both without any prior introduction to the recipe, SPICE resulted in a shorter average cooking duration (15.77\% reduction) and fewer average stops (40\% reduction), highlighting the system's ability to streamline the cooking process even though it requires two-handed interaction. This has strong implications for using TUIs and seamless integration of digital information in the immediate workspace without the use of a screen.

Participants that used SPICE without any prior introduction to the recipe self-reported much lower perceived efficiency than the smartphone users, despite their far shorter cooking time and fewer stops. This suggests that the unfamiliarity of the SPICE system for first-time users made them feel inefficient, despite their better performance.
Studies with repeated usage would be needed to confirm whether user feelings of inefficiency were due only to unfamiliarity.

Although participants that used SPICE after a smartphone self-reported higher confidence, perceived efficiency, and perceived taste than when using the smartphone, this could be because participants were exposed to the same recipe twice, rather than because of users' comparisons of the two methods. 
Further research with larger sample sizes, different recipes, and different exposure combinations would be necessary to explore nuances in user experience more comprehensively.

\section{Future Work}
Assuming that the costs of installation (e.g., tracking system, communication network) are reduced, and the technology utilized in SPICE can be adapted to any kitchen environment, it is reasonable to believe that private kitchens, as well as professional culinary settings might incorporate a CPS to streamline and improve the process of cooking, by enhancing physical objects with their digital counterparts (e.g., data, instructions). It is possible to envision a future where a kitchen can adapt to its cook and be interacted with both digitally and physically simultaneously. More modalities such as gesture and voice recognition might be an interesting addition for the end user. Finally, more advanced CV techniques to detect and deal with uncountable ingredients (e.g., flour, salt, sugar) are natural extensions of this work. The intention of the authors is to conduct further research in this direction.

\section{Conclusion}
The intersection of the digital and the physical presents challenges in creating seamless, intuitive interfaces. Through Tangible User Interfaces (TUIs), which give digital functionalities to physical objects, it becomes possible to enhance physical processes which usually require focus and two-handed interaction such as cooking. In this paper, we proposed SPICE (Smart Projection Interface for Cooking Enhancement); a prototype that provides a TUI to interact with and manipulate whilst cooking. Experiments conducted with 30 subjects assessed the efficacy of SPICE compared to traditional recipe-following methods through a comparative usability test. Results indicate that the total duration and the number of stops during the cooking process was reduced. In addition, the user confidence, taste, and difficulty metrics also improved by using the proposed system. SPICE allowed users to employ both hands actively in the cooking process while still benefiting from digital information projected in their visual working space. This study provides insights on how to blend of physical objects and digital information, which might have beneficial implications not just in cooking, but also in other high-stakes environments where physical activities and hands-on interaction predominate such as in manufacturing, medicine, and education. 

\section{Acknowledgments}
Project supported by a 2024 Leonardo Grant for Scientific Research and Cultural Creation from the BBVA Foundation. The BBVA Foundation accepts no responsibility for the opinions, statements and contents included in the project and/or the results thereof, which are entirely the responsibility of the authors.

\bibliography{references}
\bibliographystyle{IEEEtran}
\end{document}